\documentclass{article}
\usepackage{spconf,amsmath,graphicx,color,multirow,diagbox,amssymb,threeparttable,mathtools,booktabs,float}

\title{OVERLAPPED SPEECH RECOGNITION FROM A JOINTLY LEARNED MULTI-CHANNEL NEURAL SPEECH EXTRACTION AND REPRESENTATION}
\name{{\em Bo Wu$^{1}$, Meng Yu$^{2}$, Lianwu Chen$^{1}$, Chao Weng$^{2}$, Dan Su$^{1}$, and Dong Yu$^{2}$}}
\address{  $^1$Tencent AI Lab, Shenzhen, China\\
  $^2$Tencent AI Lab, Bellevue, WA, USA\\
{\small \tt \{lambowu, raymondmyu, lianwuchen, cweng, dansu, dyu\}@tencent.com}}

\begin{document}
\ninept
\maketitle
\begin{abstract}
We propose an end-to-end joint optimization framework of a multi-channel neural speech extraction and deep acoustic model without mel-filterbank (FBANK) extraction for overlapped speech recognition. First, based on a multi-channel convolutional TasNet with STFT kernel, we unify the multi-channel target speech enhancement front-end network and a convolutional, long short-term memory and fully connected deep neural network (CLDNN) based acoustic model (AM) with the FBANK extraction layer to build a hybrid neural network, which is thus jointly updated only by the recognition loss. The proposed framework achieves 28\% word error rate reduction (WERR) over a separately optimized system on AISHELL-1 and shows consistent robustness to signal to interference ratio (SIR) and angle difference between overlapping speakers. Next, a further exploration shows that the speech recognition is improved with a simplified structure by replacing the FBANK extraction layer in the joint model with a learnable feature projection. Finally, we also perform the objective measurement of speech quality on the reconstructed waveform from the enhancement network in the joint model.

\end{abstract}


\section{Introduction}
In the presence of interfering speakers, target speech is usually distorted by competing speech signals, causing degradation on speech quality and intelligibility. Such deterioration can severely affect automatic speech recognition (ASR). Despite the great progress in deep learning techniques, conventional ASR systems usually fail in these adverse conditions. Although many techniques have been developed for speech enhancement/separation and recognition under these circumstances \cite{weng2015deep, barker2015third, wang2018supervised, wu2017end}, it still remains one of the most challenging problems in ASR to date.

Existing literature typically divides the problem into enhancement/separation and recognition stages, allowing modular research to be separately conducted. One straightforward approach is to train separation
and recognition models separately in a stage-wise manner, and in a testing phase, the waveforms or features enhanced by the enhancement/separation model are then passed to the acoustic model for recognition \cite{barker2015third,du2014robust}. Several deep learning based enhancement/separation methods have been proposed, such as deep clustering \cite{hershey2016deep}, deep attractor network \cite{luo2018speaker}, permutation invariant training \cite{yu2017permutation} and TasNet \cite{luo2019conv}. Additional improvement can be obtained with microphone array signal processing which utilizes spatial information for source discrimination. For instance, an end-to-end multi-channel convolutional TasNet with short-time Fourier transform (STFT) kernel is proposed to leverage spatial cues for speech separation \cite{bahmaninezhad2019comprehensive}. However, a large gain in speech quality does not necessarily lead to improvement in speech recognition due to the speech distortion and damage in the enhancement stage.

Jointly modeling the front-end enhancement and acoustic model is therefore a desirable solution for improving speech quality and thus recognition accuracy in noisy and multi-talker environments. The work in \cite{wang2016joint} stacks up the separation network, mel-filterbank feature extraction and acoustic model to construct a deeper and larger DNN in a monaural setup. Other studies attempt to jointly train enhancement and acoustic deep models with microphone array speech input signals. For example, researchers in \cite{hoshen2015speech,sainath2016factored} design an architecture for speech acoustic modeling from raw multi-channel noisy waveform. The work in \cite{ochiai2017unified,xu2019joint} propose to jointly train a neural beamformer and acoustic model for noise robust ASR. Nevertheless, overlapped speech is not considered.

\begin{figure*}[!t]
        \centering
        \includegraphics[width=\linewidth]{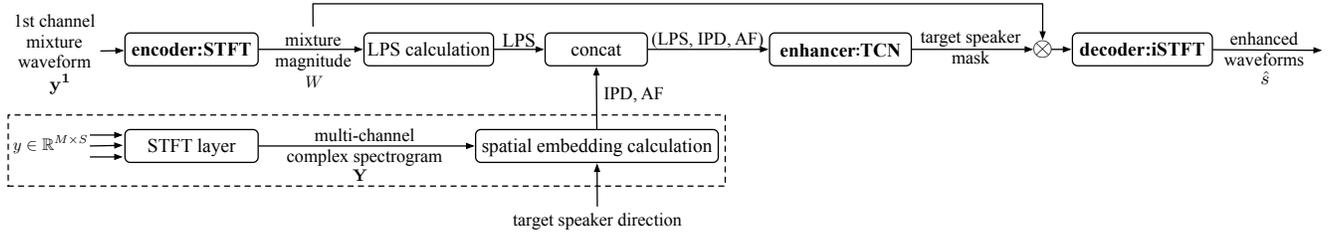}
        \caption{Block diagram of the multi-channel target speech enhancement network.}
        \label{fig:tasnet}
\end{figure*}

In this study, we aim to provide an integrated end-to-end paradigm by jointly learning multi-channel enhancement and acoustic models for overlapped speech recognition. Our contribution is two-fold: 
First, based on the multi-channel TasNet with STFT kernel \cite{bahmaninezhad2019comprehensive}, we present a joint optimization architecture of multi-channel target speech enhancement and acoustic model.
It has been shown that the model architecture with objective loss function in convolutional TasNet improves the speech separation quality significantly \cite{bahmaninezhad2019comprehensive}.
More importantly, this joint learning scheme significantly outperforms the separately optimized systems. Second, we further replace the decoder layer for waveform reconstruction and fbank layer for acoustic model feature extraction by a linear projection layer, i.e., acoustic model directly takes enhanced signal in STFT domain. This change leads to improved recognition accuracy and smaller model size, which makes it a real end-to-end multi-channel acoustic model.

The rest of this paper is organized as follows. We describe the multi-channel target speech enhancement model and CLDNN based acoustic model in Section \ref{sec:modules}. We present two joint model architectures in Section \ref{sec:jt}. Experimental results are next provided and analyzed in Section \ref{sec:exp}. We summarize this paper in Section \ref{sec:conclusion}.

\section{Multi-Channel Enhancement and Acoustic Models}
\label{sec:modules}

The proposed joint learning consists of two modules: enhancement and acoustic models, respectively. Our previous work on multi-channel convolutional TasNet with STFT kernel based target speech enhancement \cite{bahmaninezhad2019comprehensive} is adopted to recover target speaker's voice from the reverberant, noisy and multi-talker mixed signal. A CLDNN based acoustic model is utilized to predict context-independent phonemes. 

\subsection{Multi-channel target speech enhancement model}
The front-end multi-channel enhancement framework is illustrated in Fig. \ref{fig:tasnet}. It contains three major parts: an encoder (a fixed STFT convolution 1-D layer) for encoding input waveform into STFT domain, an enhancer for estimating the target speaker mask, and a decoder (a fixed iSTFT convolution 1-D layer) for waveform reconstruction. Specifically, (i) A reference channel, usually 1st channel waveform $\bf{y}^{1}$ is transformed to spectral magnitude $W$ by encoder layer with STFT kernel, which is converted to log-power spectral (LPS) feature. The LPS feature vector is then concatenated with inter-channel phase differences (IPDs) and target speaker-dependent angle feature (AF) \cite{gu2019neural}. AF feature measures the cosine distance between the steering vector of a desired direction and IPD:
\begin{equation}
\text{AF}_\theta(t, f)=\overset{K}{\underset{k=1}{\sum}}
\frac{
\mathbf{e}_{\theta,k}(f)
\frac{\mathbf{Y}_{k1}(t,f)}{\mathbf{Y}_{k2}(t,f)}}
{
\left |\mathbf{e}_{\theta,k}(f)
\frac{\mathbf{Y}_{k1}(t,f)}{\mathbf{Y}_{k2}(t,f)} \right |
}
\label{eq4}
\end{equation}
where $\mathbf{e}_{\theta,k1}(f)$ is the steering vector coefficient for target speaker from $\theta$ at frequency \emph{f}, and $\mathbf{Y}_{k1}(t,f)/\mathbf{Y}_{k2}(t,f)$ computes the IPD between microphone $k1$ and $k2$. As a result, AF indicates if a speaker from a desired direction dominates in each time-frequency bin, which drives the network to extract the target speaker from the mixture.
(ii) A temporal fully-convolutional network (TCN) \cite{luo2019conv} is adopted in the enhancement network which infers the target speaker mask. (iii) A decoder is used to reconstruct a single-channel waveform from the multiplication between mixture magnitude $W$ and target speaker mask.

Furthermore, the scale-invariant signal-to-distortion (SI-SNR) is used as the objective function to optimize the enhancement network which is defined as:
\begin{equation}
    \text{SI-SNR}:=10\log_{10}\frac
{\left\|\mathbf{s}_{\sf target}\right\|_{2}^{2}}
{\left\|\mathbf{e}_{\sf noise}\right\|_{2}^{2}}
\label{eq1}
\end{equation}
where $\mathbf{s}_{\sf target}=\left(\left<\hat{\mathbf{s}}, \mathbf{s}\right>\mathbf{s}\right)/\left\|\mathbf{s}\right\|_{2}^{2}$, $\mathbf{e}_{\sf noise}=\hat{\mathbf{s}}-\mathbf{s}_{\sf target}$, and $\hat{\mathbf{s}}$ and $\mathbf{s}$ are the estimated and reverberant target speech waveforms, respectively. The zero-mean normalization is applied to $\hat{\mathbf{s}}$ and $\mathbf{s}$ for scale invariance. We refer the
readers to \cite{bahmaninezhad2019comprehensive} for more details about the implementation of the multi-channel target speech enhancement model. 

\begin{figure}[htbp]
        \centering
        \includegraphics[width=3.5cm]{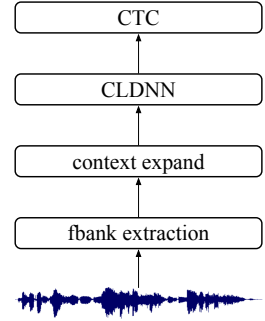}
        \caption{Block diagram of the CLDNN acoustic model.}
        \label{fig:cldnn}
\end{figure}

\subsection{CLDNN acoustic model}\label{sec:AM}

Fig. \ref{fig:cldnn} shows the adopted CLDNN acoustic model \cite{sainath2015convolutional}. The waveform-in system starts with a non-learnable feature extraction network to compute the single-channel input waveform's fbanks, and followed with an expand layer where the central frame is spliced with left and right contextual frames to leverage acoustic context information. For the architecture of CLDNN, the concatenated fbank features are first fed into two convolutional layers each of which contains 180 filters with filter size of 5. Both convolution and pooling are performed only in frequency axis. After frequency modeling with convolutional layers is performed, the outputs are passed to 4 LSTM layers for temporal modeling where each layer contains 512 hidden dimensions. Finally, we pass the outputs from LSTMs to two fully connected layers and a softmax layer to predict context-independent phonemes with the connectionist temporal classification (CTC) loss function \cite{graves2006connectionist}.

\section{End-to-End Joint Training}
\label{sec:jt}
The end-to-end design paradigm \cite{graves2006connectionist} is of growing interest in several research areas, such as speech recognition \cite{wu2017end,bahdanau2016end} and keyword spotting \cite{audhkhasi2017end}. An integrated end-to-end paradigm by jointly modeling the front-end enhancement and back-end acoustic model is therefore a desirable solution for improving 
ASR robustness in reverberant and multi-talker environments, since the target speech enhancement is directly optimized based on the speech recognition loss.

\begin{figure}[htbp]
        \centering
        \includegraphics[width=7cm]{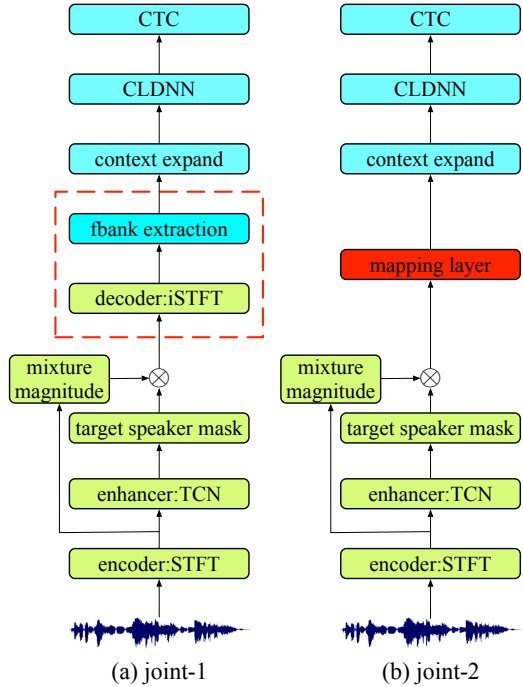}
        \caption{Block diagrams of the proposed joint models.}
        \label{fig:jt}
\end{figure}

In this paper, we adopt a hybrid deep learning framework to perform joint training of enhancement network and acoustic modeling for multi-channel overlapped speech recognition. Two integrated end-to-end paradigms are depicted in Fig.~\ref{fig:jt}. In line with a recent effort in end-to-end-modeling \cite{wu2017end}, the front-end enhancement model and acoustic model can be stacked back-to-back. Therefore, we directly stack the fbank extraction layer of the CLDNN acoustic model on top of the enhancement network's decoder layer in Fig.~\ref{fig:jt}~(a). The output layer of enhancement model becomes the input layer for acoustic modeling, which is also a hidden layer of the whole network. The same CTC object function used to train the acoustic model in Section~\ref{sec:AM} is used to fine-tune the weights of the joint model. We name this joint model as joint-1. In Fig.~\ref{fig:jt}~(b), by skipping waveform reconstruction and fbank calculation, the decoder and feature extraction in the red dash box are replaced with a linear mapping layer. The projection layer projects the 257 dimensional masked output to a 40 dimensional vector, being consistent with the conventional fbank dimension. Therefore, this architecture frees the conventional feature extraction for acoustic model and realizes a truly end-to-end joint model with trainable acoustic features which we denote as joint-2 in the following discussion.

\section{Experimental results}
\label{sec:exp}
\subsection{Dataset}
We simulated a multi-channel reverberant version of two-speaker mixture data set by AISHELL-1 corpus, which is a public data set for Mandarin speech recognition \cite{bu2017aishell}. A 6-element uniform circular array is used as the signal receiver, the radius of which is 0.035~m. The target speaker with known direction of arrival is mixed with a interfering speaker randomly at signal to interference ratio (SIR) -6, 0 or 6~dB. The classic image method \cite{ISM} is used to add multi-channel room impulse response (RIR) to each source in the mixture and reverberation time (RT60) ranges from 0.05 to 0.5~s. The room configuration (length-width-height) is randomly sampled from 3-3-2.5~m to 8-10-6~m. The microphone array and speakers are at least 0.3~m away from the wall. The distance between microphone array and speakers ranges from 1~m to 5~m. We do not apply any constraints on the direction-of-arrival differences between speakers, so that our data set contains samples with the angle difference of two speakers ranging from 0 to 180 degrees. Moreover, the train, validation and test sets consist 340, 40 and 20 speakers, respectively. There is no overlapping speakers among the three sets. All data is sampled at 16~kHz.

\subsection{Feature extraction and system setup}
\subsubsection{Multi-channel target speech enhancement model}
For the encoder and decoder settings, the kernel size and stride are 512 and 256 samples, respectively. The kernel weights are set according to STFT/iSTFT operation. 257-dimensional LPS feature is extracted based on the output of STFT kernel from the first channel mixture. 6 IPDs are extracted between microphone pairs (1, 4), (2, 5), (3, 6), (1, 2), (3, 4) and (5, 6). The target speaker's direction is assumed to be known for computing the angle feature (AF), which is feasible particularly with visual information \cite{dupont2000audio}.

\begin{table*}[!t]
\caption{WER of clean and multi-condition training of CLDNN acoustic models}
\label{table:separated}
\setlength{\tabcolsep}{4mm}
\centering
\begin{tabular}{c|c|cccc|ccc}\toprule
\multicolumn{1}{c|}{\multirow{2}{*}{\textbf{AM}}} & \multicolumn{1}{c|}{\multirow{2}{*}{\textbf{initialized}}} &  \multicolumn{4}{c|}{\textbf{training data}} & \multicolumn{3}{c}{\textbf{test data}}  \\ \cline{3-9}
\multicolumn{1}{c|}{} &\multicolumn{1}{c|}{} & \multicolumn{1}{c}{cln.}  & \multicolumn{1}{c}{rev.} & \multicolumn{1}{c}{mix-ch1}  & \multicolumn{1}{c|}{enh}  & \multicolumn{1}{c}{cln.}  & \multicolumn{1}{c}{mix-ch1}  & \multicolumn{1}{c}{enh} \\ \hline
 A1 & $\times$ & $\checkmark$ & $\times$  & $\times$ & $\times$& 11.97 & 88.70 &38.66 \\
 A2 & A1 & $\checkmark$ & $\checkmark$ & $\checkmark$  & $\times$&13.25 & 76.20 &37.40 \\
 A3 & A1 & $\times$  & $\times$ & $\times$ &$\checkmark$ & 16.76  & 91.68 & 30.69 \\
 A4 & A1 &$\checkmark$ & $\checkmark$ & $\checkmark$  &$\checkmark$ & 13.76  & 83.80 & \textbf{29.15} \\
\bottomrule
\end{tabular}
\end{table*}

\subsubsection{CLDNN acoustic model}
The architecture of CLDNN is described in Section \ref{sec:AM}. Specifically, a linear connection layer with 257-dimensional input and 40-dimensional output is used to extract fbank feature from single-channel waveforms with 25-ms window length and 10-ms hop size. After global normalization, the feature vector of the current frame is concatenated with that of the 10 preceding frames and 10 subsequent frames. The CLDNN model starts with two convolutional layers and then four LSTM layers, each with 512 hidden units, and then two full-connection linear layer plus a softmax layer. We use context-independent phonemes as the modeling units, which form 218 classes in our Chinese ASR system.

\subsubsection{ASR decoder}
A tri-gram language model (LM) is estimated on AISHELL-1 text and compiled into $G$ transducer. The whole decoding graph building procedure closely follows EESEN \cite{miao2015eesen} where $TLG$ cascade is a composition of the token transducer $T$ (used to remove the blank and repeating labels), the lexicon transducer $L$ and $G$. The ASR decoder is based on Kaldi \cite{povey2011kaldi} toolkit with minor modifications to support CTC acoustic models. The beam width is set to 18 and we use 7000 as maximum number of active tokens during decoding.

\subsection{Separated training results}
First, to obtain an optimal separated system, we train the acoustic models using different training data and evaluate them on 3 test sets in Table \ref{table:separated}. ``cln.'', ``rev.'', ``mix-ch1'' and ``enh'' denote dry clean signal of target speaker, reverberant signal of target speaker, first channel of input signal and enhanced signal, respectively. The front-end multi-channel target speech enhancement network, denoted as S0, infers a single-channel output signal ``enh'' based on 6-channel overlapped noisy speech. A competitive WER of 11.97\% is attained on clean set ``cln.'' by acoustic model A1 trained on clean set only. Multi-condition acoustic models A2, A3 and A4 are initialized with A1. For test on enhanced data ``enh'', we first obtain a WER of 38.66\% with clean model A1. Second, by adding ``rev.'' and ``mix-ch1'' into the multi-condition training set, the WER is slightly improved to 37.40\% by A2. Next, the matched acoustic model A3 trained on ``enh'' significantly decreases the WER to 30.69\%. Finally, the WER is drastically reduced from the initial 38.66\% attained with A1 down to the best one 29.15\% using A4 trained on all multi-condition sets. 29.15\% is considered as the best baseline for the comparison with joint training approaches.

\begin{table*}[htbp]
\caption{WER of joint training models}
\label{table:jt}
\setlength{\tabcolsep}{3.9mm}
\centering
\begin{tabular}{c|ccc|cccc|c}\toprule
\multicolumn{1}{c|}{\multirow{2}{*}{\textbf{system}}} & \multicolumn{3}{c|}{\textbf{SIR (dB)}} & \multicolumn{4}{c|}{\textbf{angle difference (degree)}}& \multicolumn{1}{c}{\multirow{2}{*}{\textbf{avg}}}\\ \cline{2-8} 
\multicolumn{1}{c|}{}   & \multicolumn{1}{c}{-6} & \multicolumn{1}{c}{0} & \multicolumn{1}{c|}{6}  & \multicolumn{1}{c}{0-15} & \multicolumn{1}{c}{15-45} & \multicolumn{1}{c}{45-90} & \multicolumn{1}{c|}{90-180}  \\ \hline
sept-A4  & 36.92 & 27.88 &22.74 & 38.15 & 28.81 & 27.29 & 26.19 &29.15  \\
joint-1 & 26.19 & 20.15 & 16.52 & 29.84 & 20.33 & 18.96 & 18.40 &20.94 \\
joint-2  & 26.11 & 19.57 & 16.40  & 29.77 & 20.28 & 18.63 & 18.22&20.77 \\
\bottomrule                    
\end{tabular}
\end{table*}

\subsection{Joint training results}
\begin{figure}[htbp]
        \centering
        \includegraphics[width=8cm]{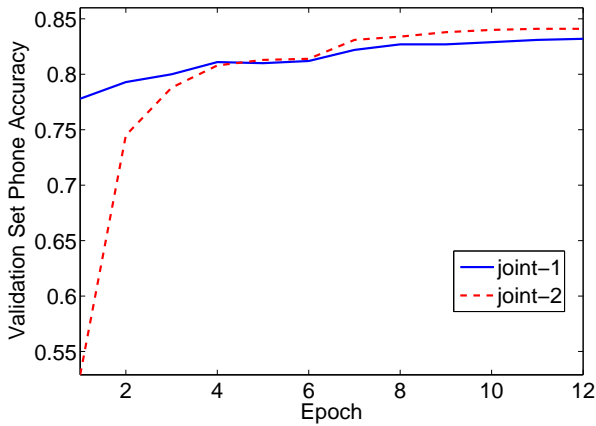}
        \caption{Validation set phone accuracy of joint training models.}
        \label{fig:acc}
\end{figure}

Next, we compare our joint models with the best separated trained system sept-A4 on data sets with different SIRs and angle differences between the two overlapping speakers in Table \ref{table:jt}. Both joint-1 and joint-2 are initialized with the above well-trained enhancement model S0 and acoustic model A4. Batch normalization is applied before context expand to accelerate convergence and improve model generalization. A significant WER decrement is achieved from 29.15\% in sept-A4 to 20.94\% using joint-1, showing a relative improvement of about 28\%. It demonstrates our proposed joint model's significant superiority. Moreover, joint-1 shows stable performances and consistently outperforms sept-A4 in all SIR and angle difference categories. More importantly, by replacing waveform reconstruction and fbank extraction in joint-1 with a linear projection layer in joint-2, WER can be further boosted to 20.77\%, even with less model parameters. It implies that the acoustic features commonly used in acoustic model training is not necessary in joint training. A detailed comparison between the proposed two joint models in terms of validation set phone accuracy is also provided in Fig. \ref{fig:acc}. It is clear that joint-2 with skipping fbank feature extraction is worse than joint-1 at the beginning epochs, but converges quickly and exceeds joint-1.

\begin{table}[htbp]
\caption{PESQ and STOI results}
\label{table:obj}
\setlength{\tabcolsep}{9mm}
\centering
\begin{tabular}{c|c|c}\toprule
\multicolumn{1}{c|}{\multirow{1}{*}{\textbf{wav}}} & \multicolumn{1}{c|}{\textbf{PESQ}} & \multicolumn{1}{c}{\textbf{STOI}} \\ \cline{1-3}
mix-ch1  & 1.85 & 0.699  \\
enh  & 2.72 & 0.882 \\
joint-1  & 2.71& 0.846 \\
joint-2  & 2.32 & 0.798 \\
\bottomrule                    
\end{tabular}
\end{table}

Finally, we present objective measures on input noisy signal, enhanced speech, and intermediate enhanced waveform outputs extracted from joint models in Table \ref{table:obj}. For joint-1, the output waveform of the decoder layer is directly used. For joint-2, although waveform reconstruction is skipped, we still can apply an external iSTFT operation on the target speaker masked output to obtain the waveform. Compared with unprocessed input channel ``mix-ch1'', front-end enhancement ``enh'' significantly boosts perceptual evaluation of speech quality (PESQ) \cite{PESQ} and short-time objective intelligibility (STOI) \cite{STOI} by 0.87 and 0.183, respectively. Moreover, although with a slight degradation of speech quality relative to ``enh'', joint-1 can obtain a relative recognition improvement of about 28\% illustrated in Table \ref{table:jt}. Furthermore, it is reasonable that enhanced output from joint-2 shows the worst score, because no waveform reconstruction constraint exists during the training. It shows that the optimal ASR loss does not necessarily lead to the best objective speech quality.

\section{Conclusions}
\label{sec:conclusion}
We propose an integrated end-to-end ASR paradigm by joint training of multi-channel target speech enhancement and CLDNN based acoustic model for overlapped speech recognition. Firstly, we directly stack these two models and jointly adjust their weights with recognition loss. This joint learning scheme achieves a WER reduction of 28\% against a separately optimized system. Next, recognition accuracy can be further improved by replacing the decoder and fbank extraction layers with a linear mapping layer, achieving an end-to-end acoustic model with trainable enhanced features.

\section{Acknowledgements}
We would like to thank our colleagues Shixiong Zhang, Jun Wang and Yong Xu for their valuable suggestions and thank the interns Rongzhi Gu, Jian Wu and Max W.Y. Lam for their code assist.

\bibliographystyle{IEEEbib}
\bibliography{strings,refs}

\begin{thebibliography}{10}

\bibitem{weng2015deep}
Chao Weng, Dong Yu, Michael~L Seltzer, and Jasha Droppo,
\newblock ``Deep neural networks for single-channel multi-talker speech
  recognition,''
\newblock {\em IEEE/ACM Transactions on Audio, Speech and Language Processing},
  vol. 23, no. 10, pp. 1670--1679, 2015.

\bibitem{barker2015third}
Jon Barker, Ricard Marxer, Emmanuel Vincent, and Shinji Watanabe,
\newblock ``{The third CHiME speech separation and recognition challenge:
  dataset, task and baselines},''
\newblock in {\em IEEE Workshop on Automatic Speech Recognition and
  Understanding (ASRU)}, 2015, pp. 504--511.

\bibitem{wang2018supervised}
DeLiang Wang and Jitong Chen,
\newblock ``Supervised speech separation based on deep learning: an overview,''
\newblock {\em IEEE/ACM Transactions on Audio, Speech, and Language
  Processing}, vol. 26, no. 10, pp. 1702--1726, 2018.

\bibitem{wu2017end}
Bo~Wu, Kehuang Li, Fengpei Ge, Zhen Huang, Minglei Yang, Sabato~Marco
  Siniscalchi, and Chin-Hui Lee,
\newblock ``An end-to-end deep learning approach to simultaneous speech
  dereverberation and acoustic modeling for robust speech recognition,''
\newblock {\em IEEE Journal of Selected Topics in Signal Processing}, vol. 11,
  no. 8, pp. 1289--1300, 2017.

\bibitem{du2014robust}
Jun Du, Qing Wang, Tian Gao, Yong Xu, Li-Rong Dai, and Chin-Hui Lee,
\newblock ``Robust speech recognition with speech enhanced deep neural
  networks,''
\newblock in {\em Proc. INTERSPEECH}, 2014.

\bibitem{hershey2016deep}
John~R Hershey, Zhuo Chen, Jonathan Le~Roux, and Shinji Watanabe,
\newblock ``Deep clustering: Discriminative embeddings for segmentation and
  separation,''
\newblock in {\em Proc. ICASSP}, 2016, pp. 31--35.

\bibitem{luo2018speaker}
Yi~Luo, Zhuo Chen, and Nima Mesgarani,
\newblock ``Speaker-independent speech separation with deep attractor
  network,''
\newblock {\em IEEE/ACM Transactions on Audio, Speech, and Language
  Processing}, vol. 26, no. 4, pp. 787--796, 2018.

\bibitem{yu2017permutation}
Dong Yu, Morten Kolb{\ae}k, Zheng-Hua Tan, and Jesper Jensen,
\newblock ``Permutation invariant training of deep models for
  speaker-independent multi-talker speech separation,''
\newblock in {\em Proc. ICASSP}, 2017, pp. 241--245.

\bibitem{luo2019conv}
Yi~Luo and Nima Mesgarani,
\newblock ``Conv-tasnet: Surpassing ideal time--frequency magnitude masking for
  speech separation,''
\newblock {\em IEEE/ACM Transactions on Audio, Speech, and Language
  Processing}, vol. 27, no. 8, pp. 1256--1266, 2019.

\bibitem{bahmaninezhad2019comprehensive}
Fahimeh Bahmaninezhad, Jian Wu, Rongzhi Gu, Shi-Xiong Zhang, Yong Xu, Meng Yu,
  and Dong Yu,
\newblock ``A comprehensive study of speech separation: spectrogram vs waveform
  separation,''
\newblock {\em arXiv preprint arXiv:1905.07497}, 2019.

\bibitem{wang2016joint}
Zhong-Qiu Wang and DeLiang Wang,
\newblock ``A joint training framework for robust automatic speech
  recognition,''
\newblock {\em IEEE/ACM Transactions on Audio, Speech, and Language
  Processing}, vol. 24, no. 4, pp. 796--806, 2016.

\bibitem{hoshen2015speech}
Yedid Hoshen, Ron~J Weiss, and Kevin~W Wilson,
\newblock ``Speech acoustic modeling from raw multichannel waveforms,''
\newblock in {\em Proc. ICASSP}, 2015, pp. 4624--4628.

\bibitem{sainath2016factored}
Tara~N Sainath, Ron~J Weiss, Kevin~W Wilson, Arun Narayanan, and Michiel
  Bacchiani,
\newblock ``Factored spatial and spectral multichannel raw waveform cldnns,''
\newblock in {\em Proc. ICASSP}, 2016, pp. 5075--5079.

\bibitem{ochiai2017unified}
Tsubasa Ochiai, Shinji Watanabe, Takaaki Hori, John~R Hershey, and Xiong Xiao,
\newblock ``Unified architecture for multichannel end-to-end speech recognition
  with neural beamforming,''
\newblock {\em IEEE Journal of Selected Topics in Signal Processing}, vol. 11,
  no. 8, pp. 1274--1288, 2017.

\bibitem{xu2019joint}
Yong Xu, Chao Weng, Like Hui, Jianming Liu, Meng Yu, Dan Su, and Dong Yu,
\newblock ``Joint training of complex ratio mask based beamformer and acoustic
  model for noise robust asr,''
\newblock in {\em Proc. ICASSP}, 2019, pp. 6745--6749.

\bibitem{gu2019neural}
Rongzhi Gu, Lianwu Chen, Shi-Xiong Zhang, Jimeng Zheng, Yong Xu, Meng Yu, Dan
  Su, Yuexian Zou, and Dong Yu,
\newblock ``Neural spatial filter: target speaker speech separation assisted
  with directional information,''
\newblock in {\em Proc. INTERSPEECH}, 2019, pp. 4290--4294.

\bibitem{sainath2015convolutional}
Tara~N Sainath, Oriol Vinyals, Andrew Senior, and Ha{\c{s}}im Sak,
\newblock ``Convolutional, long short-term memory, fully connected deep neural
  networks,''
\newblock in {\em Proc. ICASSP}, 2015, pp. 4580--4584.

\bibitem{graves2006connectionist}
Alex Graves, Santiago Fern{\'a}ndez, Faustino Gomez, and J{\"u}rgen
  Schmidhuber,
\newblock ``Connectionist temporal classification: labelling unsegmented
  sequence data with recurrent neural networks,''
\newblock in {\em Proc. ICML}, 2006, pp. 369--376.

\bibitem{bahdanau2016end}
Dzmitry Bahdanau, Jan Chorowski, Dmitriy Serdyuk, Philemon Brakel, and Yoshua
  Bengio,
\newblock ``End-to-end attention-based large vocabulary speech recognition,''
\newblock in {\em Proc. ICASSP}, 2016, pp. 4945--4949.

\bibitem{audhkhasi2017end}
Kartik Audhkhasi, Andrew Rosenberg, Abhinav Sethy, Bhuvana Ramabhadran, and
  Brian Kingsbury,
\newblock ``End-to-end asr-free keyword search from speech,''
\newblock {\em IEEE Journal of Selected Topics in Signal Processing}, vol. 11,
  no. 8, pp. 1351--1359, 2017.

\bibitem{bu2017aishell}
Hui Bu, Jiayu Du, Xingyu Na, Bengu Wu, and Hao Zheng,
\newblock ``{AISHELL-1: An open-source mandarin speech corpus and a speech
  recognition baseline},''
\newblock in {\em Proc. O-COCOSDA}, 2017, pp. 1--5.

\bibitem{ISM}
E.A. Lehmann and A.M. Johansson,
\newblock ``Prediction of energy decay in room impulse responses simulated with
  an image-source model,''
\newblock {\em The Journal of the Acoustical Society of America}, vol. 124, no.
  1, pp. 269--277, 2008.

\bibitem{dupont2000audio}
St{\'e}phane Dupont and Juergen Luettin,
\newblock ``Audio-visual speech modeling for continuous speech recognition,''
\newblock {\em IEEE Transactions on Multimedia}, vol. 2, no. 3, pp. 141--151,
  2000.

\bibitem{miao2015eesen}
Yajie Miao, Mohammad Gowayyed, and Florian Metze,
\newblock ``Eesen: End-to-end speech recognition using deep rnn models and
  wfst-based decoding,''
\newblock in {\em Proc. ASRU}, 2015, pp. 167--174.

\bibitem{povey2011kaldi}
Daniel Povey, Arnab Ghoshal, Gilles Boulianne, Lukas Burget, Ondrej Glembek,
  Nagendra Goel, Mirko Hannemann, Petr Motlicek, Yanmin Qian, Petr Schwarz,
  et~al.,
\newblock ``The kaldi speech recognition toolkit,''
\newblock in {\em Proc. ASRU}, 2011.

\bibitem{PESQ}
{ITU-T, Rec. P.862},
\newblock ``{Perceptual evaluation of speech quality (PESQ): An objective
  method for end-to-end speech quality assessment of narrow-band telephone
  networks and speech codecs},''
\newblock {\em Int. Telecommun. Union-Telecommun. Stand. Sector}, 2001.

\bibitem{STOI}
Cees~H Taal, Richard~C Hendriks, Richard Heusdens, and Jesper Jensen,
\newblock ``An algorithm for intelligibility prediction of time--frequency
  weighted noisy speech,''
\newblock {\em IEEE Transactions on Audio, Speech and Language Processing},
  vol. 19, no. 7, pp. 2125--2136, 2011.

\end{thebibliography}
\end{document}